\documentclass[12pt]{article}  
\newcommand{\B}{Ba\v zant}  

                            \newcommand{\no}{\noindent}
 \newcommand{\bc}{\begin{center}}
 \newcommand{\ec}{\end{center}}
                   \newcommand{\bfr}{\begin{flushright}}
                   \newcommand{\efr}{\end{flushright}}
   \newcommand{\ii}{\item}
     \newcommand{\be}{\begin{enumerate}}
     \newcommand{\ee}{\end{enumerate}}
        \newcommand{\bi}{\begin{itemize}}
        \newcommand{\ei}{\end{itemize}}
            \newcommand{\bd}{\begin{description}}
            \newcommand{\ed}{\end{description}}
                \newcommand{\beq}{\begin{equation}}
                \newcommand{\eeq}{\end{equation}}
                  \newcommand{\bea}{\begin{eqnarray}}
                  \newcommand{\eea}{\end{eqnarray}}
        
  \newcommand{\bu}{$\bullet$ \hspace*{.5em}}

      \newcommand{\bfi}{\begin{figure}}
      \newcommand{\efi}{\end{figure}}
\newcommand{\bay}{\begin{array}{l}}
\newcommand{\eay}{\end{array}}
            \newcommand{\dd}{\mbox{d}}

    \newcommand{\pa}{\partial}
    \newcommand{\del}{\delta}
    \newcommand{\Del}{\Delta}
    
    \newcommand{\la}{\lambda}
    
    \newcommand{\al}{\alpha}

\usepackage{graphicx}
\usepackage{amsmath}
\usepackage{breqn}
\usepackage{textcomp}
\usepackage{amssymb}
\begin{document}   
\thispagestyle{empty}
        \hspace*{1mm}  \vspace*{-0mm}
\noindent {\footnotesize {{\em
\hfill      Submitted to arxiv, July 3, 2016} }}
\vskip 1.5in
\begin{center}
{\Large {\bf  Direct Testing of Gradual PostPeak Softening of Notched Specimens of Fiber Composites Stabilized by Enhanced Stiffness and Mass
           }}\\[9mm]

{\large {\sc Zden\v ek P. Ba\v zant, Viet T. Chau, Gianluca Cusatis, Marco Salviato}}
\\[1in]

{\sf Report No. 16-07/d}\\[1.3in]

Department of Civil and Environmental Engineering
\\ Northwestern University
\\ Evanston, Illinois 60208, USA
\\[1in]  {\bf July 3, 2016}    
\end{center}

\clearpage   \pagestyle{plain} \setcounter{page}{1}

\begin{center}
 {\Large {\sf Direct Testing of Gradual PostPeak Softening of Notched Specimens of Fiber Composites Stabilized by Enhanced Stiffness and Mass
        }} \\[7mm]  {\large {\sf
        Zden\v ek P. Ba\v zant\footnote{
Dept. of Civil and Env. Engineering, Northwestern University, 2145 Sheridan Road, CEE/A135, Evanston, Illinois 60208;
z-bazant@northwestern.edu.},
        Viet T. Chau\footnote{
Dept. of Civil and Env. Engineering, Northwestern University, 2145 Sheridan Road, Evanston, Illinois 60208;
VietChau2017@u.northwestern.edu}
        Gianluca Cusatis\footnote{
Dept. of Civil and Env. Engineering, Northwestern University, 2145 Sheridan Road, Evanston, Illinois 60208;
g-cusatis@northwestern.edu} and
        Marco Salviato\footnote{
William E. Boeing Dept. of Aeronautics and Astronautics, University of Washington, Guggenheim Hall, Seattle, Washington 98185; formerly Northwestern University}
  }}\\
\end{center} \vskip 5mm   

\no {\bf Abstract:}\ {\sf

Static and dynamic analysis of the fracture tests of fiber composites in  hydraulically servo-controlled testing machines currently in use shows that their grips are much too soft and light for observing the postpeak softening. Based on static and dynamic analysis of the test setup, far stiffer and heavier grips are proposed. Tests of compact-tension fracture specimens of woven carbon-epoxy laminates prove this theoretical conclusion. Sufficiently stiff grips allow observation of a stable postpeak, even under load-point displacement control. Dynamic stability analysis further indicates that stable postpeak can be observed under CMOD control provided that a large mass is rigidly attached to the current soft grips. The fracture energy deduced from the area under the measured complete load-deflection curve with stable postpeak agrees closely with the fracture energy deduced from the size effect tests of the same composite. Previous suspicions of dynamic snapback in the testing of composites are dispelled. So is the previous view that fracture mechanics was inapplicable to the fiber-polymer composites.
 }

\subsection*{1. Introduction}

The material failure criteria for fiber-polymer composites have for a long time been expressed in terms of stresses or strains. Examples are the maximum stress, maximum strain, deviatoric strain energy, and tensor polynomial criteria \cite{TsaWu72}. Their general applicability, however, is an in-grained antiquated myth, surviving from pre-computer age. Such criteria apply to plastic materials, which exhibit no strain localization instability, no material characteristic length and no deterministic size effect.

In reality, fiber composites are quasibrittle materials (which also include concrete---as the archetypical case, tough ceramics, rocks, sea ice, rigid foams, bone, etc.). All quasibrittle materials fail by localization of softening damage into a discrete fracture. In contrast to plasticity, they exhibit a material characteristic length which inevitably leads to a strong energetic (or non-statistical) size effect when geometrically similar structures of different sizes are compared \cite{BazCed91,BazPla98,BazDanLi96}.

Two basic types of size effect must be distinguished. Here the focus is on the Type 2 size effect, which occurs when a large notch or stress-free crack exists at maximum load. This size effect is weak for small specimens not much larger than the periodicity of the weave or the size of the representative volume element (RVE), for which it may seem that the stress or strain failure criteria work. But with increasing structure size, there is a gradual transition to the strong size effect of fracture mechanics caused by stored energy release associated with stress redistribution during damage. It may be noted that the Type 1 size effect occurs in structures that fail right at the initiations of a macro-crack from a damaged RVE, and represents a combination of deterministic and statistical (or Weibull) size effects (omission of the deterministic aspect led to an erroneous conclusion [e.g.]\cite{GreWisHal07}, namely that the Weibull modulus is a geometry-dependent variable rather than a material constant).

At mesh refinement, the use of stress or strain criteria inevitably causes a loss of objectivity and convergence problems \cite{BazPla98}. For this reason, as well as fundamentally, realistic failure analysis must be based on quasibrittle fracture mechanics, which evolved since its dawn in the mid 1970s into a mature and widely accepted theory. Fracture mechanics, in fact the classical linear elastic fracture mechanics (LEFM), is well accepted for delamination fracture of layered two-dimensional (unstitched) fiber-composite laminates, and there is even an ASTM test to determine the corresponding fracture energy \cite{ASTM5528}.

The fact that quasibrittle fracture mechanics must apply to in-plane or flexural loading of fiber composite laminates was demonstrated by the  numerous size effect tests performed, beginning in 1996 \cite{BazDanLi96,Baz99,BazKimDan99,BazZhoNov04,SalKirAsh16} on geometrically similar notched specimens. However, to many engineers and researchers the size effect tests have been unconvincing, for two reasons: 1) some of them erroneously considered the size effect to be statistical, due to material randomness (although this is possible only for Type 1 failures); 2) others objected that a gradual postpeak softening could never be observed in experiments. The specimens always failed explosively right after attaining the maximum load, and the load applied by the testing machine dropped suddenly to zero. The sudden drop seemed to indicate a LEFM behavior, but the LEFM clearly did not fit test data.

\subsection*{Highly Stiff Test Frame with Fast Servo-Control via CMOD}

A similar history occurred long ago for concrete and rock. Until the 1960's it was believed that concrete and rock explode at maximum load and the load applied by the testing machine drops suddenly to zero. Then in the 1960s, several researchers, including Hughes, Chapman, Hillsdorf, R\" usch, Evans and Marathe \cite{RusHil63, HugCha66, EvaMar68, HeiHilFin69} came up with the idea of using, for both tensile and compressive tests of concrete, a much stiffer loading frame and fast hydraulic servo-control. Suddenly, a gradual postpeak decline of the compressive or tensile load could be observed. Efforts to stabilize postpeak in compression testing of rock were made, beginning 1963, with the work of Neville G.W. Cook and Charles Fairhurst at University of Minnesota \cite{WawFai70, HudBroFai71, Fai16}. The stability of postpeak was further enhanced by controlling the test electronically with a gage measuring the crack-mouth opening displacement (CMOD). A servo-controlled stiff machine of MTS Corporation was built in 1967. 

This discovery opened a revolution in the mechanics of concrete and rock, and was one essential factor that prompted the development of quasibrittle fracture mechanics. The stabilizing effect of machine stiffness was mathematically demonstrated by static stability analysis in \cite{Baz76} (see also \cite{BazCed91}), which led to an equation for the machine stiffness as a function of the maximum steepness of the postpeak load-deflection curve.

Unfortunately, the same measures did not work for fiber composites. The same stiff frames with fast servo-control did not suffice. The CMOD control of notched compact tension specimens and of edge-notched strips was tried at Northwestern, but did not work. Neither did the control of crack tip opening displacement (CTOD). The reason will be clarified here.

\subsection*{Static Criterion of Stability}

For the purpose of static analysis, test system can be considered as the coupling of two elements: 1) the testing machine frame of stiffness $K_m$ with the specimen grips (or fixtures) of stiffness $K_g$ form one elastics $K_m$, and 2) the test specimen, of tangential (or incremental) stiffness $K_s$ is the second element. Figure \ref{f1} shows a schematic of the testing frame.

Based on the second law of thermodynamics, the test setup becomes unstable if the there exists a perturbing load that produces negative work on the test setup, which causes an increase of entropy (in detail see Chapter 13 of \cite{BazCed91}). So, we imagine a perturbing load $\del P$ to be applied axially at the load-point of the test specimen. The combined stiffness of the machine with the grips is $K_{mg} = 1/(1/K_m + 1/K_g)$ and the total stiffness $K_t$ resisting $\Del P$ is $K_{mg} + K_s$. The displacement under $\Del P$ in the direction of $\del P$ is $\del v = \del P/K_t$.

The system is stable if and only if $\del v / \del P$ is $ > 0$. So the static stability condition is $K_t > 0$ or
 \beq \label{x1}
   K_t\ =\ \frac 1 {1/K_m + 1/K_g}\ + K_s\ >\ 0
 \eeq

Consider now a typical 20 ton testing machine (e.g., MTS) used in the testing of composites and a typical compact tension specimen consisting of 24 layers of woven carbon fiber epoxy composites for a total thickness of 5.4 mm (see Figure \ref{f2}). The typical characteristics are:
 \bd \setlength{\itemsep}{-1.5mm}
\ii \bu Machine frame stiffness: $K_m$ = 260 MN/m;
\ii \bu Stiffness of standard specimen grips: $K_g$ = 0.768 MN/m;
\ii \bu The steepest slope of the measured postpeak load-deflection curve of the test specimen: $K_s = - 0.830$ MN/m;
\ii \bu Mass of machine frame: $m_m$ = 500 kg;
\ii \bu Mass of grips (fixture):$m_g$ = 0.919 kg;
\ii \bu Characteristic halftime of machine hydraulics delay: $\tau$ = 0.02 s (defined as the time to approach halfway to a suddenly changed setting of electronic control).
\ed
Using these value, we find from Eq. (\ref{x1}) that
 \beq  \label{x2}
  K_t = - 0.06426 \mbox{MN/m} < 0~~~~ ...~\mbox{\bf unstable}
 \eeq
i.e., the tested specimen is unstable. This agrees with the common wisdom of the last 50 years---it is impossible to observe gradual postpeak softening in tensile tests of fiber composites. But is it, really?

The grip stiffness is an aspect that has so far eluded attention. It has generally been assumed that the standard grips provided by the manufacturers are stiff enough. But comparison of $K_g$ with $K_m$ suggests otherwise. It was, therefore, decided to produce special massive grips about 10-times heavier (Fig. \ref{f3}), with the following stiffness and mass;
 \bd \setlength{\itemsep}{-1.5mm}
\ii \bu $\tilde K_g$ = 192.4 MN/m~~~(= 74\% $K_m$)
\ii \bu $\tilde m_g$ = 9.419 kg~~~  (= 10 $m_g$)
 \ed
Eq. (\ref{x1}) gives:
 \beq  \label{x3}
  K_t = 109.6 MN/m > 0~~~~ ...~\mbox{\bf stable}
 \eeq
Eureka, stability is achieved.

Using these grips, the composite compact tension specimens (Fig. \ref{f3}) exhibited stable progressive softening, and not only under CMOD control but also under load-point control. In fact, this was not unexpected because the fracture energy $G_f$ deduced from the size effect tests was larger that the maximum possible area under a snapback curve. In fact, the area indicated by $G_f$ required progressive softening with a finite negative slope, and this is what is now observed.

From Eq. (\ref{x1}), one can obtain the minimum necessary stiffness of the grips:
 \beq \label{xx1}
  K_g\ =\ \frac 1 { 1/|K_s| - 1 / K_m }~~~~~(K_s < 0)
 \eeq

\subsection*{Test Stabilization by CMOD or CTOD Control}

The usual way to stabilize postpeak softening is to control the Crack Mouth Opening Displacement (CMOD), $w$. Let $w$ be the relative displacement across the crack mouth or across the crack-tip region, and denote as $F$ a fictitious force working on $w$ althogh in reality $F$ must vanish. The cross compliance $C_{sc}$ between the load point and displacement $w$ is derived by considering the relations
 \bea \label{e1}
  \dd u &=& C_s \dd P + C_{sc} \dd F
 \\ \label{e2}
  \dd w &=&  C_{sc} \dd P + C_{cc} \dd F
 \eea
where $P$ = applied force; $C_s = 1/K_s$ = direct (load-point) compliance of the specimen. According to the LEFM \cite[e.g.]{BazPla98},
 \beq 
  C_{sc} = \int_0^{\al} k(\al') k_c(\al') \dd \al'
 \eeq
where $\al = a/D$, $a$ = crack length, $D$ = specimen dimension; $k(\al),  k_c(\al)$ = dimensionless stress intensity factors due to applied load $P$ and to fictitious load $F$; $k = K_I b \sqrt{D}/E$ where $K_I$ = actual stress intensity factor; and $D, b$ = specimen width and dimension. Because $F=0$, Eqs. (\ref{e1})--(\ref{e2}) reduce to: $\dd u = C_s \dd P$ and $\dd w = C_{sw} P$.

The advantage of using CMOD or CTOD control is that, during the fracture test, $w$ always increases. So, by controlling $w$, the postpeak softening can be measured even if the specimen is unstable under load-point control. But there is a caveat---the response of the hydraulics must be fast enough. For notched concrete and rock specimens it has been fast enough. But for strong and very light specimens such those of woven laminates it as apparently not been fast enough, since all attempts to measure the postpeak by means of CMOD control have failed. To explain, we turn to dynamic analysis of stability.

\subsection*{Using Mass to Make Hydraulics Response Delay Tolerable}

For half a century, the impossibility to obtain a stable postpeak softening under CMOD or CTOD control been been blamed on some unspecified peculiar property of composites. However, the present discovery of stable postpeak with very stiff grips shows that there is nothing peculiar in the material behavior of composites. So the only possible explanation is that the response of hydraulics is not fast enough (even with the optimal PID setting). To check it, let us now conduct dynamic analysis of test stability.

The test set-up may be idealized as shown in Fig. \ref{f1}, where $u(t)$ is the load-point displacement of the test specimen; $v(t)$ is the displacement at the attachment of the grips (or fixture) to the loading frame, and $y(t)$ is the input from the electronic control, representing the prescribed load-displacement history ($t$ = time). The effective mass of the machine frame is denoted as $m_m$, and the mass of the grips as $m_g$, including the mass of the specimen, which is, however, negligible in the case of composites). The mass and stiffness of the load cell are considered to be included in $M$ and $K_M$.

To control the test, the controller of the machine sends a signal to the servo-valve. The hydraulic pressure on the piston increases and the piston moves, but not immediately. The halftime, $\tau$, of the hydraulics response delay, $\tau$, which is of the order of 0.02 s (and is assumed to correspond to the optimized PID setting), may be modeled by a damper of viscosity constant
 \beq \label{tau}
  \eta\ =\ K_m\, \tau
 \eeq

Because the system can be considered incrementally linear, it will suffice to analyze the response to a sudden unit change of $y$, i.e., $y = H(t)$ where $H$ denotes the Heaviside step function. Because only infinitely small increments are considered, the response may be considered to be linear and $K_s$ to be constant, characterizing the steepest postpeak slope of the postpeak load displacement curve ($K_s < 0$). We also assume that no unloading would occur (because, for unloading, $K_s$ would switch to a positive value).

The equations of motion can be derived from the Lagrange equations:
\bea \label{e6a}
  &&\frac{\pa}{\pa t} \left(\frac{\pa L}{\pa \dot v} \right) - \frac{\pa L}{\pa v}
  + \frac{\pa {\cal D}}{\pa \dot v} = 0
  \\ \label{e6b}
  &&\frac{\pa}{\pa t} \left(\frac{\pa L}{\pa \dot u} \right) - \frac{\pa L}{\pa u}
  + \frac{\pa {\cal D}}{\pa \dot u} = 0
 \\ \label{e7a}   \mbox{where}~~~
   &&L = T-V
   \\ \label{e7b}
   &&T = \mbox{$\frac 1 2$}\, M \dot v^2 + \mbox{$\frac 1 2$}\, m_g \dot u^2\\
   &&V = \mbox{$\frac 1 2$} K_M (v-y)^2 + \mbox{$\frac 1 2$}\, K_g (v-u)^2 + \mbox{$\frac 1 2$}\, K_s u^2~~~~~~~~~~~~~
 \\ \label{e8a}
   && {\cal D} = \mbox{$\frac 1 2$}\, K_M\, \tau (\dot v - \dot y)^2
 \eea
where the superior dots denote derivatives with respect to time $t$. These equations yield the following equations of motion:
 \bea \label{e1}
  M \ddot v + K_M (v-y) + K_M\, \tau (\dot{v}-\dot{y}) + K_g (v-u) &=& 0~~~~~
 \\ \label{e2}
  m_g \ddot u + K_g (u-v) + K_s u &=& 0
 \eea
where $K_M\, \tau$ was substituted. It is convenient to rewrite the equations of motion in the phase space by introducing new variables:
\beq \label{e5}
  x_1 = u,~ x_2 = \dot{u},~ x_3 = v,~ x_4 = \dot{v}
 \eeq
Substitution into the equations of motion gives a system of first-order ordinary linear differential equations in matrix form:
 \beq  \label{matr eq}
  \left\{ \begin{array}{c}
    \dot{x_1}        \\
    \dot{x_2}        \\
    \dot{x_3}        \\
    \dot{x_4}
  \end{array} \right\}
    =
  \left[ \begin{array}{cccc}
    0 & 1 & 0 & 0  \\
    -\frac{K_g+K_s}{m_g} & 0 & -\frac{K_g}{m_g} & 0 \\
    0 & 0 & 0  & 1 \\3e3w
    \frac{K_g}{M} & 0 & -\frac{K_M+K_g}{M} & -\frac{K_M\, \tau} M
  \end{array} \right]
  \left\{\begin{array}{c}
    {x_1}        \\
    {x_2}        \\
    {x_3}        \\
    {x_4}
  \end{array} \right\}
    +
  \left\{ \begin{array}{c}
    0        \\
    0        \\
    0        \\
    \frac{K_M} M y + \frac{K_M\, \tau} M \dot{y}
  \end{array} \right\}
 \eeq
The homogeneous part of this first-order matrix differential equation is satisfied by functions of the form $x_n = a_n e^{\la t}$ ($n = 1,2,3,4$). Substitution into the homogeneous part of the foregoing matrix differential equation yields a homogeneous matrix algebraic equation for the column matrix of $a_n$. It has a nonzero solution if and only if $\la$ is equal to the eigenvalues of the square matrix in Eq. (\ref{matr eq}). The solution is stable if and only if, for all the eigenvalues,
 \beq
  \mbox{Re} (\la)\ <\ 0
 \eeq


For calculations, we consider first the aforemention machine and test properties with the standard (light) grips. Then the following column matrix of eigenvalues is calculated:
 \beq \label{eigen1}
  \{ \lambda\ \} =\
   \left\{  \begin{array}{c}
    80.65           \\
   -1.035 \cdot 10^4\\
   -77.17           \\
   -53.87
   \end{array} \right\}
 \eeq
The presence of a positive eigenvalue, $\la_1$, indicates that, with the normal grips, the test of postpeak is unstable, which means that postpeak softening cannot be observed, as known from experience. Also known that the eigenvalues are real, which means that detection of static instability in Eq. (\ref{x2}) was sufficient.

Second, consider the new grips of mass $m_g = 10 m_{g0}$ = 9.419 kg and stiffness $K_g = 10 K_{g0} = 2.6 \cdot 10^9$ N/m  (Fig. \ref{f3}). All the other parameters remain the same. Calculations yield the eigenvalue matrix:
 \beq \label{eigen2}
  \{ \lambda\ \} =\
   \left\{  \begin{array}{c}
   -15.78 + 4.52 i \cdot 10^3\\
   -15.78 - 4.52 i \cdot 10^3\\
   -1.032 \cdot 10^4		  \\
   -50.129
   \end{array} \right\}
 \eeq
All the eigenvalues are negative. So the specimen is stable, even under load-point control. This confirms the previous finding by static stability analysis, Eq. (\ref{x3}).

\subsection*{Added Mass as a Way to Stabilize CMOD Control of Tests with Soft Grips}

Why hasn't the CMOD or CTOD control worked with the standard grips? Obviously, it would have to work if the response of the controls were infinitely fast. But with hydraulic system this is impossible. The specimen accelerates fast in dynamic motion before the hydraulics can adjust the displacement. Obviously, the way to slow down the acceleration is to attach mass to the grip. So, consider the grip mass $m_g$ to be increased to $\bar m_g = 10 m_g, 100 m_g and 1000 m_g$. This leads to the sets of eigenvalues listed under the diagrams of Fig. \ref{f4}. All of these sets include an eigenvalue with Re($\la) > 0$, which means the specimens are unstable---but unstable under controlled load point displacement.

With a view of CMOD control, lets calculate the response of the system for input $y(t) = H(t)$ under four initial conditions $u = \dot u = v = \dot v = 0$. The solution is obtained as a sum of the particular solution and a linear combination of four eigenvectors. The response curves of $u(t)$ are plotted in the four diagrams of Fig. (\ref{f4}). Note that, with increasing grip mass, the time at the onset of sharp exponential  acceleration of displacement $u(t)$ (briefly `onset time') greatly increases.

To compare the onset time with the performance of the hydraulics, we plot in the figure vertical lines at time of 0.1 s, which is 5-times longer than the halftime of the hydraulics delay. We assume that by this time the CMOD control should be able to enforce the specified load-point displacement $u$ with sufficient accuracy.

In the first diagram of Fig. \ref{f4}, which corresponds to the standard (light) grips, the rise of exponential acceleration of $u(t)$  begins much before the critical time of 0.01 s. Obviously, the controls are too slow to prevent this acceleration which inevitably leads to sudden failure. However, as seen in the third diagram, the grip mass of $100 m_g$ postpones the acceleration well beyond 0.1 s, and here the hydraulics controlled by the CMOD should evidently be able to impose the required load-point displacement. According to the second diagram for $10 m_g$, it seems the exponential acceleration could also be prevented, but better informed analysis of the hydraulic system and trial testing may be needed. And the fourth diagram, for 1000 $m_g$, is obviously an unnecessary overkill.

\subsection*{Experimental Verification of Stable Postpeak Softening of a Compact Tension Specimen}

Fig. \ref{f3} shows the compact-tension fracture specimen of woven carbon-epoxy specimens used to study the postpeak behavior and determine the fracture energy $G_f$ of the material and the general view of the test setup in a MTS testing machine.


Fig. \ref{f6} shows the photos of the current standard grips (on the left) and of the proposed massive grips (on the right) that successfully stabilized the postpeak.

Fig. \ref{f7} demonstrates several stable postpeak load-delection diagrams measured with the proposed stiff grips on the compact-tension specimens.

\subsection*{Agreement of $G_f$ from Size Effect Tests with $G_f$ from the Area under Complete Load-Displacement Curve}

The area $A$ under the complete stable load-displacement curve of the fracture specimen allows determining the fracture energy, $G_f$, of the material; $G_f = A / L b$ where $l$ = length of the broken ligament and $b$ = specimen thickness (provided that energy dissipation outside the fracture is negligible.

Another, easier, way to determine the fracture energy is the size effect method \cite{BazPla98}. For this purpose it was more convenient to use
tensile tests of geometrically similar edge-notched strip specimens of 3 different sizes, as shown in Fig. \ref{f8}. This method uses only the maximum loads and the postpeak softening is not needed. So it sufficed to conduct the tests with the standard grips, even though they failed right after the peak load. The fracture energies obtained from the size effect and from the postpeak were respectively:
 \beq \label{GfGf}
  \mbox{from size effect:}~~~G_f = 73.7 ~\mbox{N/m~~~and~~~
  from postpeak:}~~~G_f - 78 ~\mbox{N/m}
 \eeq
It is remarkable that the difference between these two values is only 5.8 \%.

\subsection*{Nonexistence of Snapback and of Conflict with Size Effect}

Because it was impossible to observe postpeak softening it has been believed for decades that the fracture specimens of composites exhibit a severe snapback. Thanks to stiffening the grips, we now see that this view was incorrect.

The misconception of snapback further shed false doubts on the applicability of fracture mechanics to fiber composites. The area under the supposed snapback curve, which must in any case be smaller than the area under a load-deflection curve with sudden vertical load drop from the peak-load point, gave a fracture energy much smaller than that deduced from the size effect, or from the measured drop of complementary energy of test specimen.

For some investigators, this severe mismatch was another reason to consider quasibrittle cohesive fracture mechanics inapplicable to fiber composites. Now we see that this interpretation was mistaken.

The present analytical and experimental results, and especially Eq. (\ref{GfGf}), prove that fracture mechanics is perfectly applicable to fiber composites.

\subsection*{Conclusions}

 \be \setlength{\itemsep}{-1.5mm} \item
The specimen grips (or fixtures) of the hydraulic servo-controlled testing machines do not have sufficient stiffness to enable measurement of stable postpeak softening.
 \ii
By stability analysis based on the second law of thermodynamics, it is  shown that the cause of pervasive failure to measure postpeak softening during decades of prior testing of composites was the instability due to insufficient stiffness of the specimen grips.
 \ii
Based on static stability analysis of the test setup, it is proposed to use grips that are stiffer by about two orders of magnitude. Calculations show that, in this way, stability is achieved. Experiments on compact-tension fracture specimens of composites confirm it, even for the case of load-point displacement control.
 \ii
Calculations of the eigenvalues of the equations of motion of the test setup confirm the static stability analysis. They also indicate an enormous effect of the increase of mass of the grips on the development of dynamic instability. It is shown that, if a sufficient mass is attached to the existing soft grips, the postpeak response under CMOD control can be stabilized.
 \ii
The present stability analysis also explains why the switch in the 1960s to far stiffer testing frames sufficed to stabilize the postpeak softening in concrete and rocks. The specimens of those materials are far more massive and their grips or attachments have naturally been far stiffer than those used for fiber composites.
 \ii
The previous view that the impossibility of measuring postpeak softening implied a severe snapback is not correct. There is no snapback. The inference that the smallness of the area under the supposed snapback curve conflicts with $G_f$ measured from the size effect or from the energy release was thought to invalidate fracture mechanics. This inference is false. The present results prove that fracture mechanics is perfectly applicable to fiber composites.
 \ii
The present results also prove that the previously widespread use of plasticity-based failure envelopes in terms of stress or strain has not been correct.
 \ee
\section*{Acknowledgments}
\small
Partial funding obtained from NSF under Grant CMMI-1439960, from ARO under Grant W911NF-15-1-0240 and from USCAR (through Ford Motor Co.) under Grant 13-2856-AMR/DE-EF0005661, all to Northwestern University, is gratefully acknowledged.

\linespread{1}\selectfont




\clearpage

\section*{Appendix: Details of Experiments}

\subsection*{Materials}
Experiments were conducted on woven composite specimens manufactured by compression molding. A DGEBA-based epoxy resin was chosen as polymer matrix whereas the reinforcement was provided by a twill 2x2 fabric made of carbon fibers. The material was characterized following ASTM standard procedures \cite{ASTM5045} testing under compact tension. The material used in this testing is a $[0^{\circ}]_8$ lay-up with a constant thickness of
approximately 1.8 mm.

In addition to woven composite, the experiments were also conducted on Fiberglass Reinforced Polyester (FRP) composite. The thickness of this material is about 10 mm.


\subsection*{Specimen characteristics}
The modified Compact Tension (CT) specimen geometry was recently developed at Northwestern University research group to produce stable crack growth in a specimen so that the composite damage zone could be investigated. Initially, a 2 mm width notch was created by using diamond band saw. Then, the notch was extended by using artistic wire saw in order to create a shape notch tip of 0.2 mm in radius. The CT specimen geometry with a sharp notch tip is stable under displacement control and is large enough so that the boundaries do not greatly affect the damage zone size or shape.\\

The specimen was loaded in tension through pins located above and below the notch. Generally, these holes can't be drilled using steel drill bits because carbon fiber is harder than the steel from which the drill is made. In order to avoid damage due to fiber tear-out and delamination around the holes from using steel drill bits, abrasive cutting is used to create these holes. Holes for the pin loading were drilled using tungsten grinding bits for nonmetals.\\

To prevent buckling due to thin thickness of the woven composite material, the final specimen thickness was 5.4 mm. \\

Further, intra-laminar size effect tests were conducted on single-edge-notched tension (SENT) specimens (see Figure \ref{f8}), using a [$0^{\circ}$]$_8$ lay-up with a constant thickness of approximately 1.9 mm. The SENT specimens were preferred to Double-Edge Notched Tension (DENT) specimens, in which there is response path bifurcation such that only one of the two cracks can propagate, causing asymmetric response \cite{BazTab92}.

Specimens of three sizes (three for each size), geometrically scaled in two-dimension (see Table \ref{T2}) as 1:2:4, were tested. The first half of the notch was made by means of a diamond coated bend saw which provided a width of roughly 1 mm whereas the second half was made using a diamond-coated miniature blade thanks to which a width of 0.2 mm was obtained in all cases. Accordingly, the resulting crack tip radius was 0.1 mm, about 70 times smaller than the size of a Representative Unit Cell (RUC) of the material.

All the specimens were prepared with 38 mm long glass/epoxy tabs for gripping purposes. The tab length (grip constraint) was not scaled because it has no appreciable effect on the stored energy and because fracture always occurs away from the grips.

The top surface of all the SENT specimens investigated was treated to allow Digital Image Correlation (DIC) analysis. A thin layer of white paint was deposited on a $D \times D$ area embedding the crack. Then, black speckles of average size 0.01 mm were spray-painted on the surface after drying.

\subsection*{Testing}
The Compact Tension (CT) test set-up is shown in Fig. 2a. A universal testing machine was used to load the CT specimens at a rate of 1 mm/min. Tensile loading was applied through
20 mm diameter pins inserted through the holes shown in Fig. 1a.
An extensometer was attached to the specimens
to measure the pin opening displacement (POD). The load cell signal and the extensometer signal were output and recorded. \\

\subsection*{Experimental results}
\subsubsection*{Size Effect Test Results}
After the completion of the experiments, the load and displacement data were analyzed.
Figure \ref{f9} shows, for the various sizes, the typical load-displacement plots reported. It is
worth noting that, for the largest specimen size, these curves are almost linear up to failure,
which is an indication of pronounced brittle behavior.

After reaching the peak load, the specimens exhibited snap-back instability for all investigated sizes. As a consequence, the failures were catastrophic (dynamic), and occurred shortly after the peak load. Damage consisting of microcracks in layers, delamination between layers before peak load and tow breakage and pull-out was observed in the tests.

It is worth noting that, according to strength-based criteria (such as e.g. Tsai and Wu \cite{TsaWu72} among others), the nominal strength does not depend on the structural size. However, Table 3 does show a significant decrease of $\sigma_N$ with increasing characteristic size of the specimen. It is clear that strength based criteria cannot capture this trend. From size effect results, the initial fracture energy can be calculated: one has $G_f$ = 73.7 N/mm

\subsubsection*{Compact Tension Test Results}
Figure \ref{f7} gives the load vs. pin opening displacement (POD) curves of all the tested specimens.
For woven composite specimens, it can be seen that, the load vs. POD curves are approximately linear up to the first load drop. After that the crack progresses in a series of small jumps which result in further load drops.

In contrast, the FRP specimens showed a large degree of non-linearity in the load vs. POD curves. Figs. \ref{f7}a,b and \ref{f7}c,d show typical failure modes of woven composite and FRP composite specimens, respectively. The failure of woven composite is characterized by development of fiber breakage along the center line of the specimen toward the back. A small amount of splitting can be observed either side of the fiber breaks which has resulted in the paint for the Digital Image Correlation (DIC) system flaking away from the surface.

From the load vs. POD curves (Fig. \ref{f7}), the fracture energy of the composites can be calculated using the following equation \cite{BazPla98}:
 \beq \label{e1}
  G_f = \frac{W}{b \cdot l_l }
 \eeq
 where $G_f$ = fracture energy, $W$ = area under the load displacement curve, $b$ = thickness of the specimen and $l_l$ = ligament length.

 The calculated fracture energies for woven composites are 76.34 N/mm and 79.74 N/mm for specimen 1 and specimen 2 respectively. These results are very close to the calculated fracture energy using the size effect test.

 Also, the fracture energies for FRP composites are 14.16 N/mm and 13.09 N/mm for specimen 1 and specimen 2 respectively.

\clearpage
\listoftables
\listoffigures
\clearpage
\begin{table}[ht]
\centering  
\begin{tabular}{l c c c c c}
\hline
\rule{0pt}{4ex}Size & Width, &Gauge length,  &Length, &Crack length,  &Thickness, \\
&$D$&$L$&$L=L+2L_t$&$a_0$&$t$\\[1 ex]
\hline
Small & $20$ & 44.5 & 120.5 & 4 & 1.9\\
Medium & $40$& 89.0 & 165.0 & 8 & 1.9\\
Large & $80$ & 178.0& 254.0 & 16& 1.9 \\
\hline
\multicolumn{6}{l}{Units: mm. Tab length $L_t=$ 38 mm for all investigated sizes.}
\end{tabular}
\caption{\sf Geometrical specifications of the SENT specimens under study}
\label{T2}
\end{table}
%
%
%

\bfi \center
\includegraphics[trim=6cm 0cm 6cm 0cm, clip=true,clip=true,width = 1\textwidth]{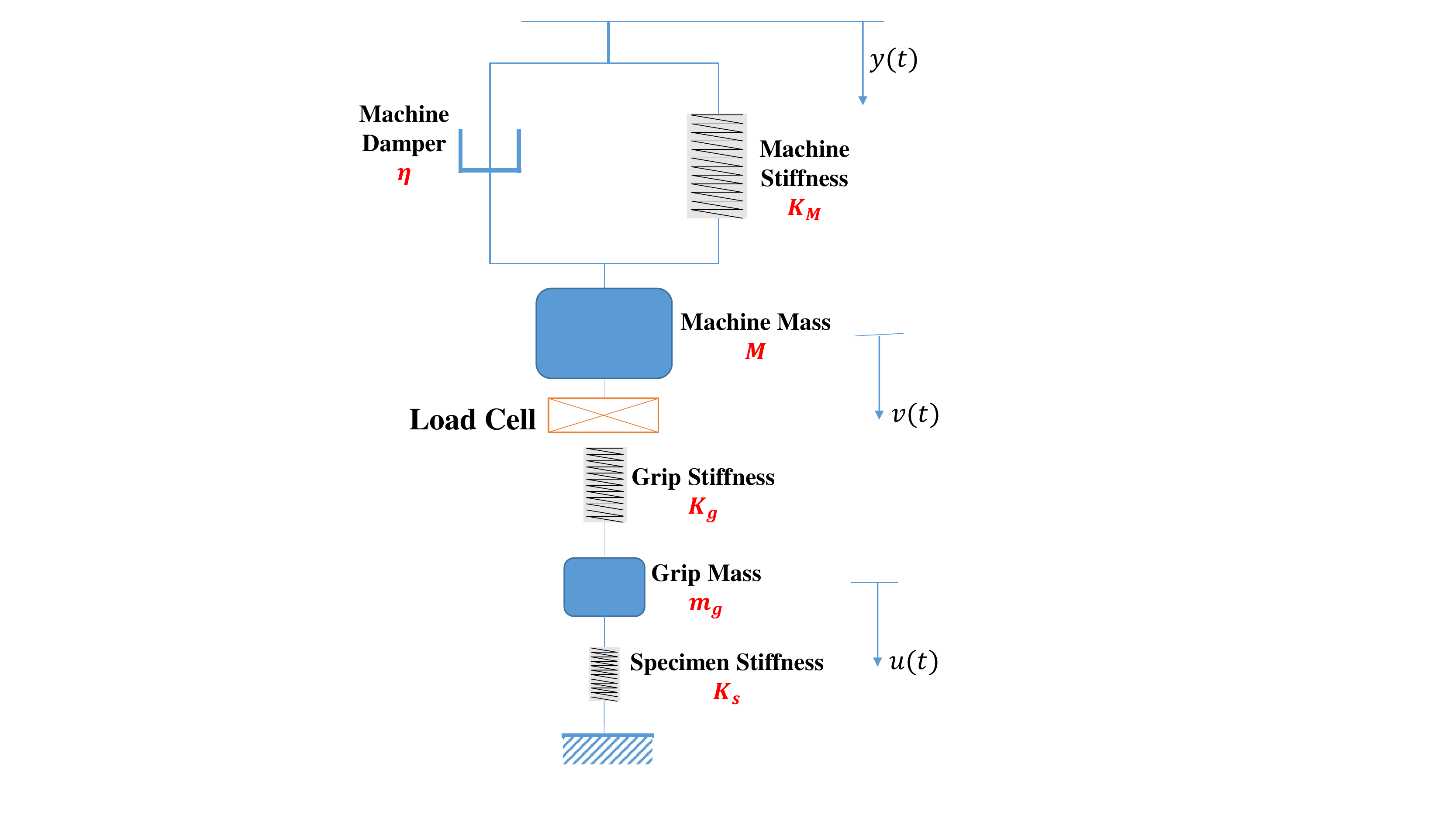}   \caption{\label{f1} \sf Simplified Schematic of a universal servo-hydraulic frame} \efi
\clearpage

%
\bfi \center
 \includegraphics[clip=true,width = 1.0\textwidth]{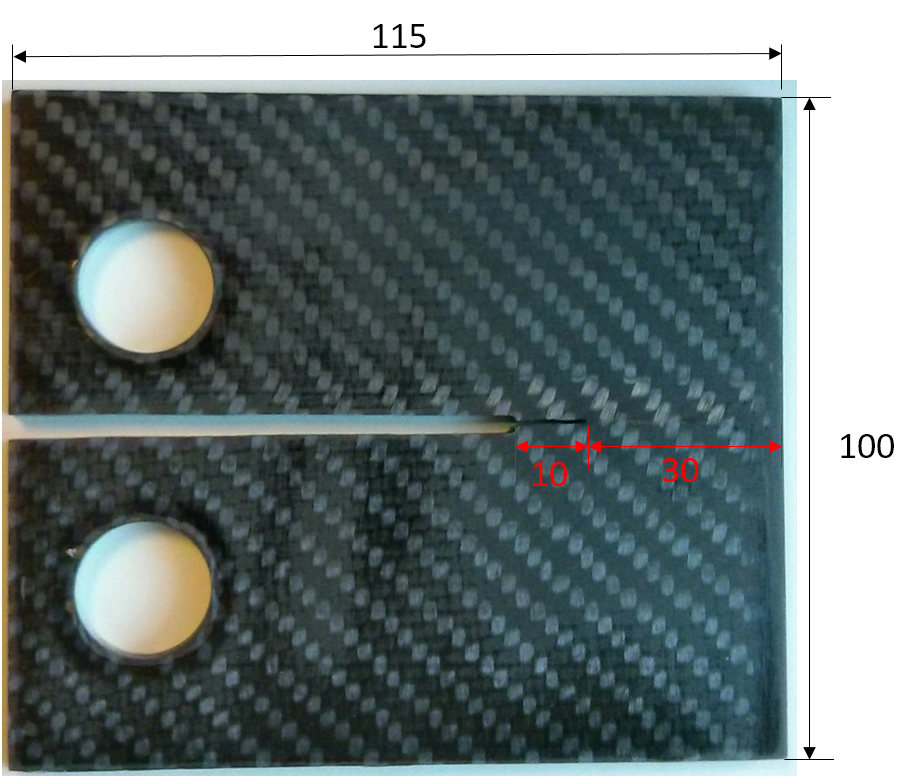}  \caption{\label{f2} \sf 2D Woven Composite Specimen Dimensions} \efi

\clearpage
%

\bfi \center
  \includegraphics[trim=3cm 0cm 3cm 0cm, clip=true,clip=true,width = 1\textwidth]{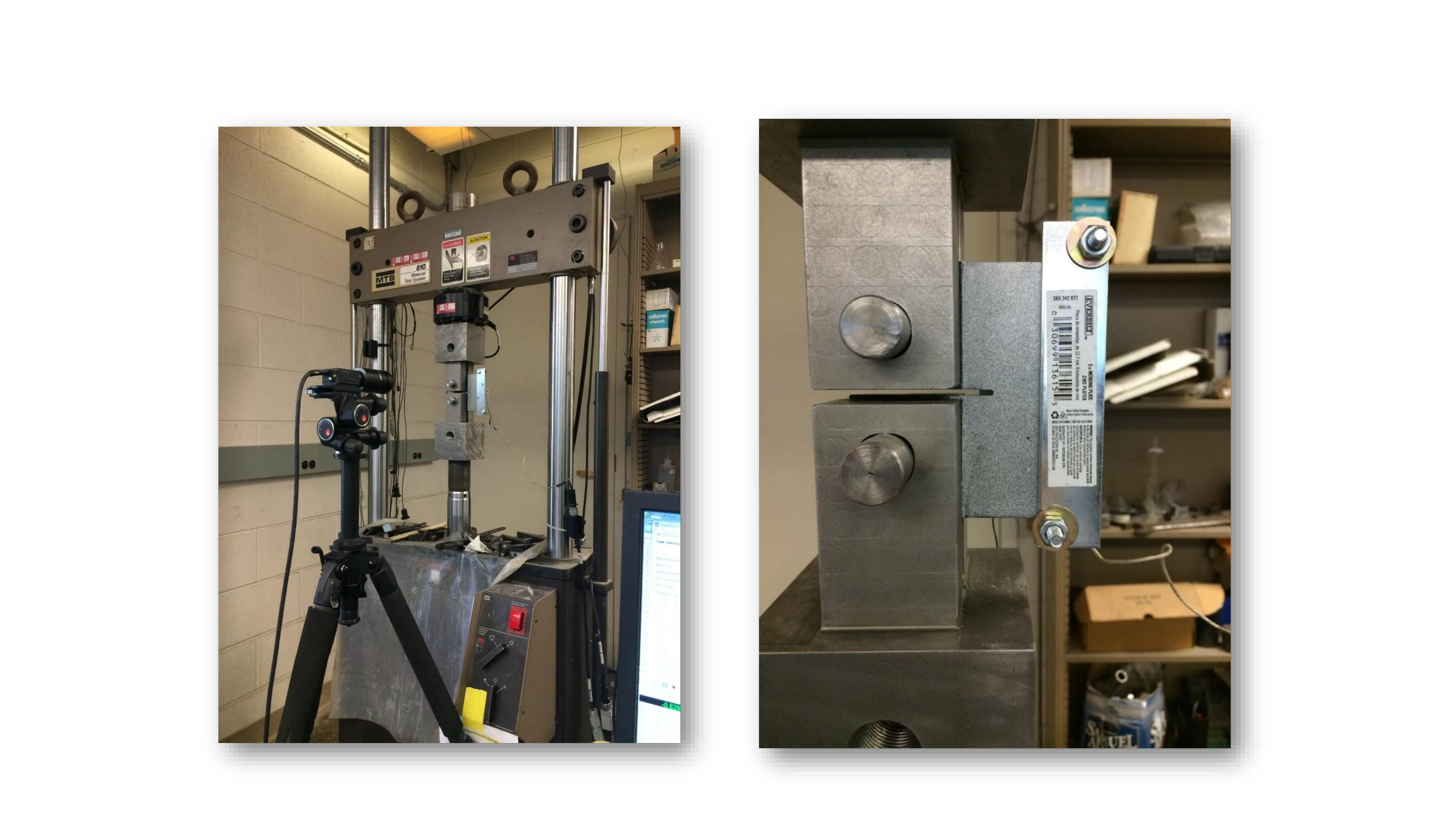} \caption{\label{f3} \sf a) Experimental setup considered in the analysis and b) massive grips designed for the tests } \efi

\clearpage

\bfi \center
  \includegraphics[trim=4cm 0cm 4cm 0cm, clip=true,clip=true,width = 1\textwidth]{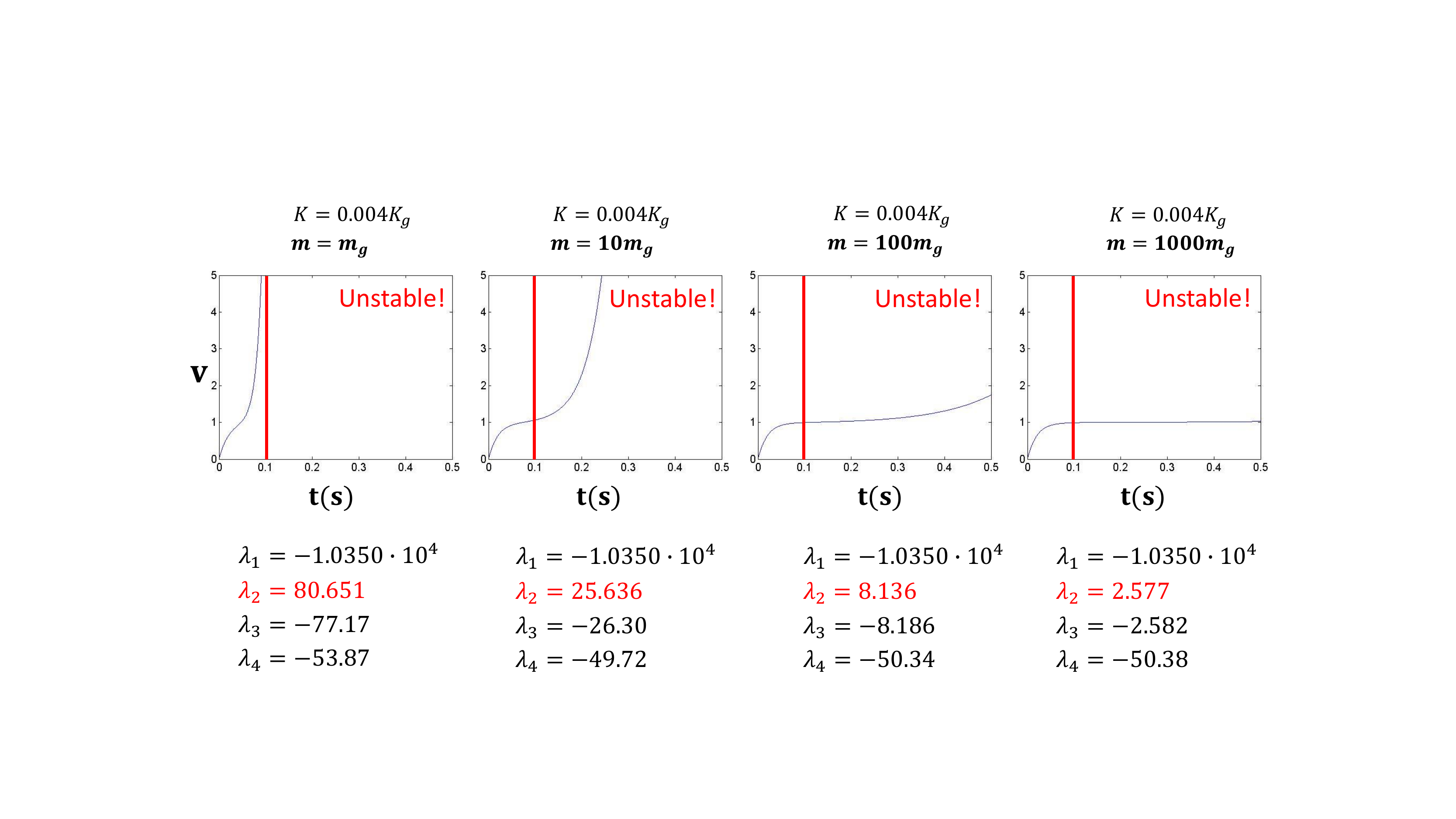} \caption{\label{f4} \sf Load-point control stability analysis for various values of grip mass, m_g } \efi
%

\bfi \center
\includegraphics[trim=3cm 0cm 3cm 0cm, clip=true,clip=true,width = 1\textwidth]{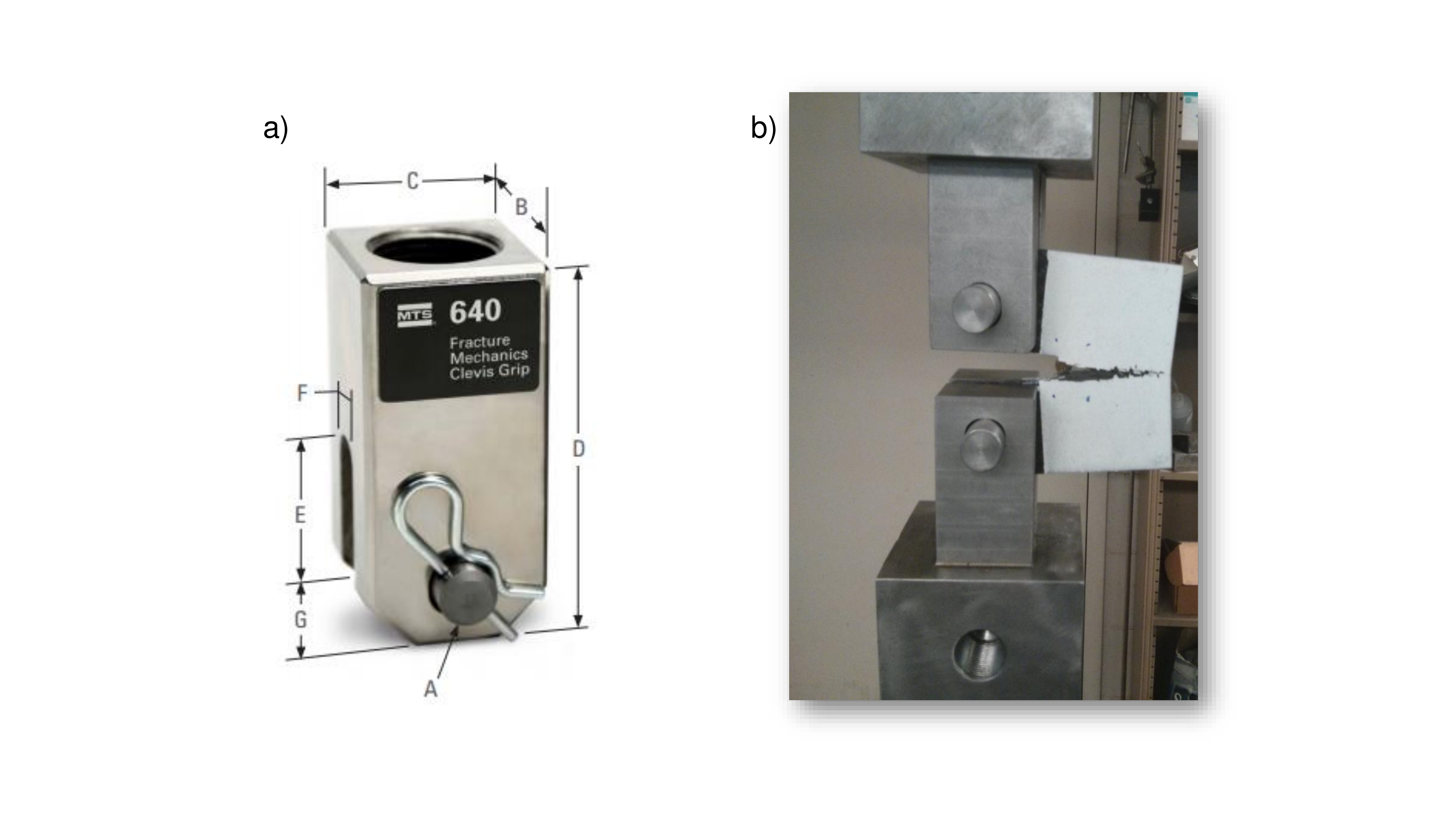}   \caption{\label{f6} \sf a) Example of MTS Standard Grip and b) newly designed grips for higher stiffness and mass} \efi
\clearpage

%
\bfi \center
\includegraphics[trim=3cm 0cm 3cm 0cm, clip=true,clip=true,width = 1\textwidth]{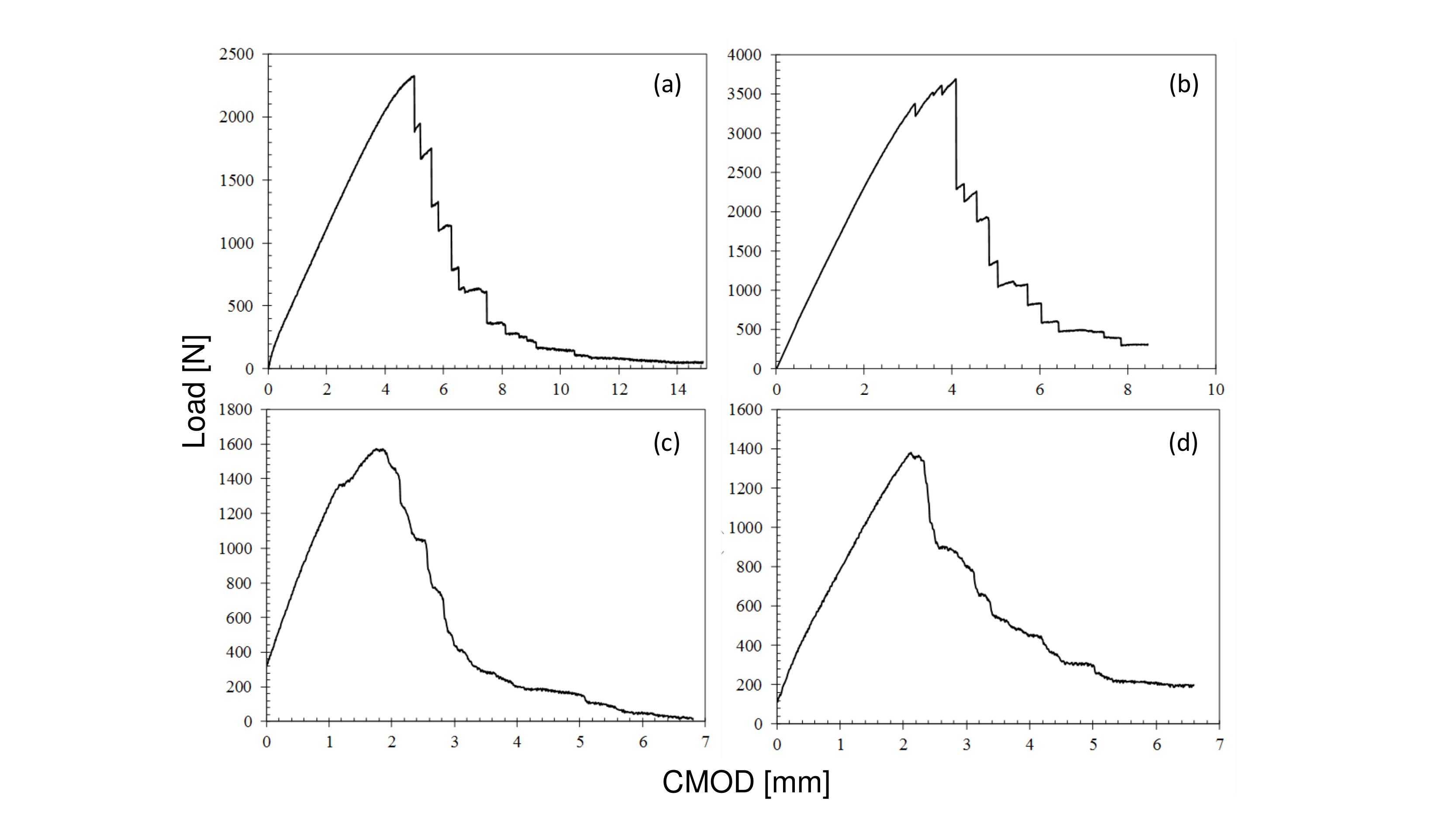}  \caption{\label{f7} \sf Typical stable load-displacement curves obtained with the newly designed grips for carbon fiber woven composites (a) and (b) and for glass fiber textile composites (c) and (d).} \efi
\clearpage
%
%
\bfi \center
  \includegraphics[trim=0cm 0.9cm 0.5cm 1cm, clip=true,clip=true,width = 1.0\textwidth]{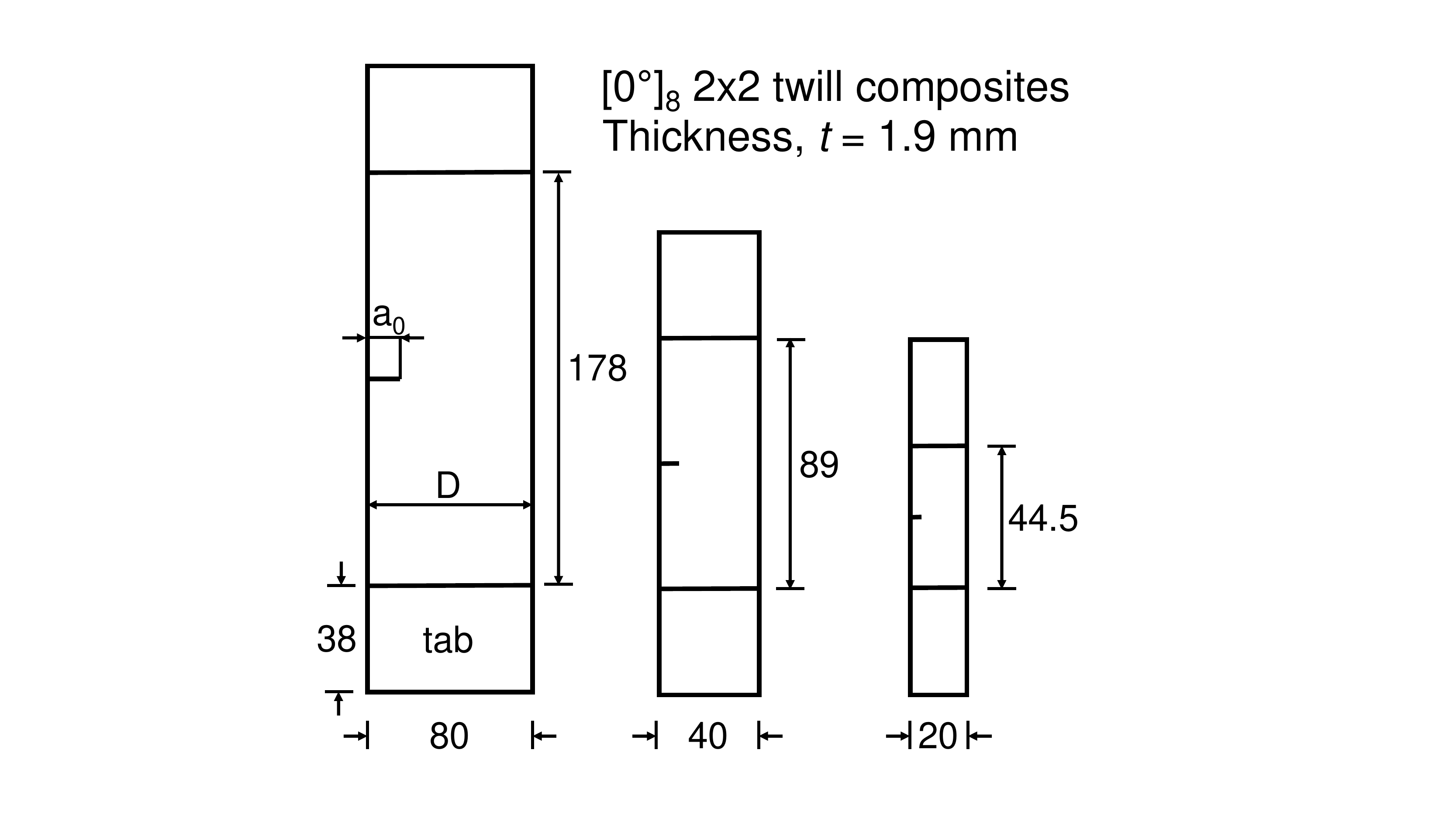} \caption{\label{f8} \sf Geometry of Single Edge Notch Tension (SENT) specimens under study. Units: mm.} \efi
\clearpage
\bfi \center
  \includegraphics[trim=1.0cm 3.5cm 0.5cm 3cm, clip=true,clip=true,width = 1\textwidth]{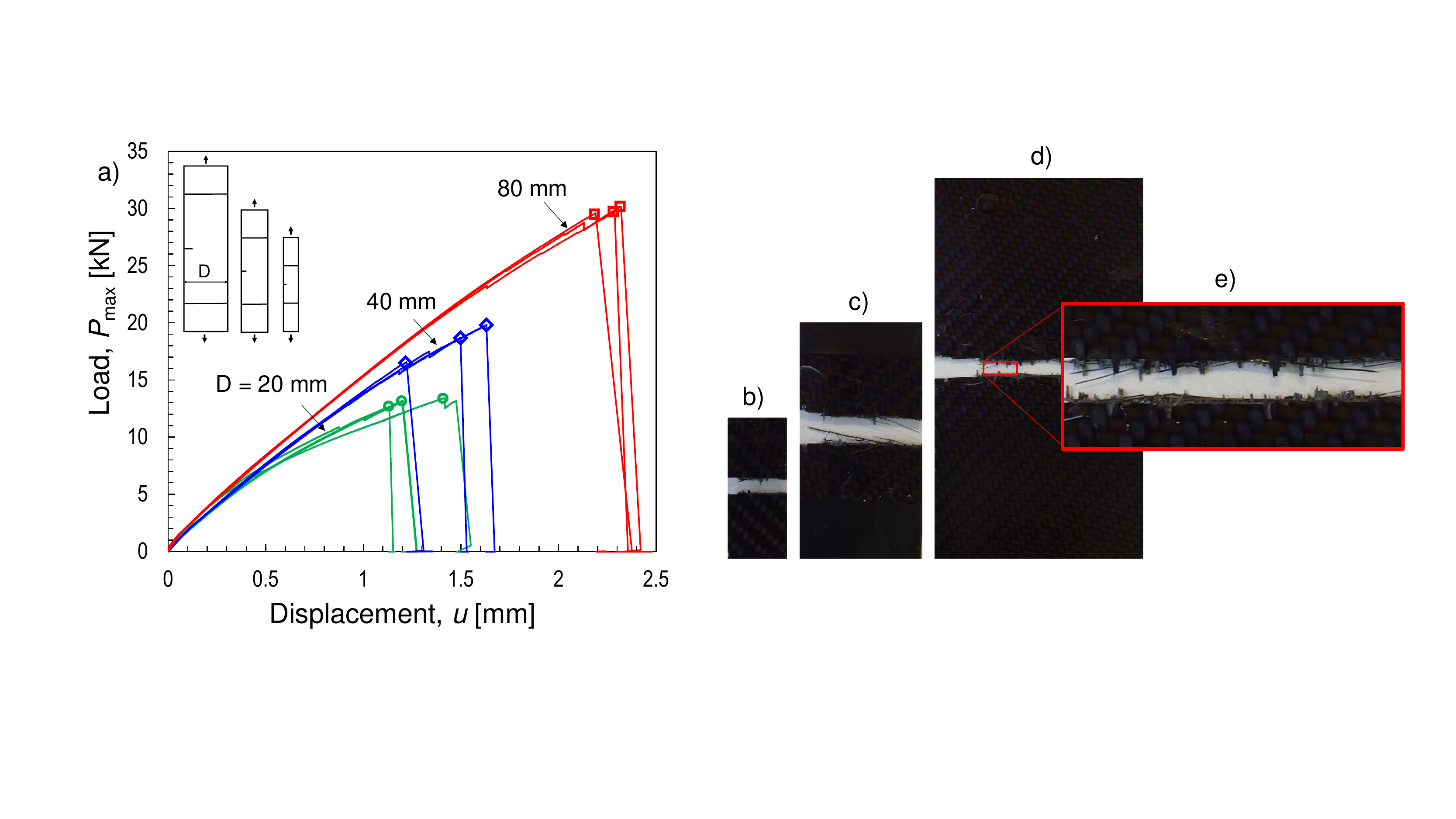} \caption{\label{f9} \sf a) Typical load-displacement curves of $[0^{\circ}]_8$ geometrically-scaled SENT specimens of various sizes, showing decreasing nonlinearity increasing specimen dimensions. Typical failure patterns of Single Edge Notched specimens for width b) $D=20$ mm, c) $D=40$ mm and d) $D=80$ mm. e) Magnification of fracture surface for the large size specimen showing extensive tow failure and pull-out.} \efi
%

  \end{document}